\documentclass[]{aa}

\usepackage{psfig,graphicx}

\begin{document}

\title{AM~1934-563: A giant spiral polar-ring galaxy in a triplet\thanks{Based 
on observations made at the Observat\'{o}rio do Pico dos
Dias (OPD), operated by the MCT/Laborat\'{o}rio Nacional de 
Astrof\'{i}sica,Brazil}}

\author{V. Reshetnikov\inst{1,}\inst{2,}\inst{4}, F. Bournaud\inst{2},
F. Combes\inst{2},
M. Fa\'{u}ndez-Abans\inst{3}, \and M. de Oliveira-Abans\inst{3}}

\offprints{resh@astro.spbu.ru }

\institute{Astronomical Institute of St.~Petersburg State 
University, 198504 St.~Petersburg, Russia 
\and
Observatoire de Paris, LERMA, 61 Av. de l'Observatoire, 75014 Paris, France
\and
MCT/Laborat\'{o}rio Nacional de Astrof\'{i}sica, Caixa Postal 21,
CEP:37.504-364, Itajub\'{a}, MG, Brasil
\and 
Isaac Newton Institute of Chile, St.~Petersburg Branch}

\date{Received  / Accepted 15 September 2005}

\titlerunning{AM~1934-563}

\authorrunning{Reshetnikov et al.}

\abstract{
We have observed the emission-line kinematics and photometry
of a southern triplet of galaxies. The triplet contains a
giant spiral galaxy AM~1934-563 which optical structure resembles 
a polar-ring galaxy: distorted spiral disk, seen almost edge-on,
and a faint large-scale (45 kpc in diameter) warped
structure, inclined by 60$^{\rm o}$--70$^{\rm o}$ with respect
to the disk major axis. The triplet shows relatively small
velocity dispersion (69 km~s$^{-1}$) and a large crossing time (0.17
in units of the Hubble time). The disk of AM~1934-563 demonstrates
optical colors typical for an early-type spirals, strong
radial color gradient, and almost exponential surface
brightness distribution with an exponential scale-length value 
of 3.1 kpc ($R$ passband). The galaxy shows a maximum rotation velocity
of about 200 km~s$^{-1}$ and it lies close to the Tully-Fisher relation 
for spiral galaxies.
The suspected polar ring is faint ($\mu(B) \geq 24$) and strongly 
warped. Its total luminosity comprises (10-15)\% of the total 
luminosity of AM~1934-563. We then try to model this system using numerical 
simulations, and study its possible formation mechanisms. We find that 
the most robust model, that reproduces the observed characteristics of 
the ring and the host galaxy, is the tidal transfer of mass from a 
massive gas-rich donor galaxy to the polar ring. The physical properties 
of the triplet of galaxies are in agreement with this scenario.
}

\maketitle
\section{Introduction}

Polar-ring galaxies (PRG) are peculiar objects in which a polar or 
very inclined ring surround the host galaxy. They teach us a lot on 
galaxy formation. At least two main scenarios are invoked to form 
these systems: either the major merger scenario, with a head-on 
merging of two galaxies with their disks oriented perpendicularly 
(e.g., Bekki 1998), or the accretion scenario, where a gas-rich 
donor loses matter to form the ring (e.g., Schweizer et al. 1983, 
Reshetnikov \& Sotnikova 1997). The accretion scenario itself could 
either correspond to a small companion being disrupted during a minor 
merger, or a major encounter with tidal mass transfer from a massive 
donor to the host (Bournaud \& Combes 2003, hereafter BC03). In 
previous simulations, BC03 have shown that the accretion scenario 
offers more chances to form the observed systems, and that a 
discriminating characteristic can be the existence, in the merger scenario, 
of a diffuse stellar background around the polar-ring system. In the accretion 
scenario, the matter is not as much dispersed spatially. It is important to 
test on several well-observed polar-ring systems the possible scenarios, and 
the present system offers a good example.
 
The southern peculiar galaxy AM~1934-563 was classified by
Whitmore et al. (1990) (PRC, polar rings catalogue) as a good candidate 
for polar-ring galaxies. (The object name is PRC~B-18 according to the PRC.)
The faint extended feature crosses the main galaxy (disturbed
edge-on galaxy with inclined dust lane) at an angle of about
60$^{\rm o}$ from the major axis (Figs.~1 and 2). Both components --
the main body and the suspected ring -- are slightly S-shaped.

AM~1934-563 has two nearby companions of similar size and
magnitude (Figs.~1 and 2). These are PGC~400092 (NW companion) 
and PGC~399718 (S companion), both of unknown redshift.   
Other galaxies of comparable magnitudes are located about
10$'$ N of AM~1934-563 and belong to the AM~1934-562 group
of galaxies. 

AM~1934-563 is an almost unexplored galaxy. Reshetnikov et al. (2001)
have found fast rotation of the gas in the central region of
the galaxy, as far as signs of Sy2 or LINER activity. 
van Driel et al. (2002) have detected the 21-cm HI line emission
toward AM~1934-563 but the observed HI profile can be confused
with another galaxy of the triplet. 

We present here new photometric and spectroscopic data for
AM~1934-563 and two companion galaxies. The new data 
reveal that the three galaxies form a physical triplet. Global observational 
structure and kinematics of AM~1934-563 allow to conclude 
that this galaxy is a real polar-ring galaxy with a spiral
host and inclined ring/disk structure.

We also model our observations using N-body numerical simulations. We 
study different formation scenarios for this polar ring, and find that 
only one is likely to reproduce the observed properties of AM~1934-563. 
We then conclude that this polar ring has been formed by tidal accretion 
of material from a gas-rich galaxy, that may still be observed today in 
the triplet. 
Let us stress that AM~1934-563, with its massive spiral host, 
is peculiar among PRGs, which most frequently have a small early-type hosts 
(often S0s). Thus, we study the formation of AM~1934-563 in particular, 
and do not pretend that our results will strictly extent to PRGs in general, 
even if many PRGs have probably formed through the same accretion mechanism (see BC03).

The observations that have been made are presented in Sect.~2. Their 
results are analyzed in Sect.~3. In Sect.~4, we model this system, 
and discuss its formation mechanism.

\section{Observations and reductions}

\subsection{Photometric observations}

The photometric observations were performed with the 1.6-m
telescope at the OPD (MCT/LNA), Brasil on August 2002. 
The telescope was equipped with direct imaging camera \#1 and
a thick, back-illuminated 2048$\times$2048, 13.5-$\mu$m square
pixels CCD detector \#98 (pixel size is 0\farcs18$\times$0\farcs18).
The readout noise was 2.4$e^-$ and the gain, 2.5$e^-$/ADU.

The data were acquired with standard Johnson $B$, $V$ and Cousins
$R$, $I$ filters. Photometric calibration was accomplished using repeated
observations of standard stars from the Landolt (1983) and 
Graham (1982) lists. The seeing during the observations was 1\farcs3. 
A log of observations is given in Table~1. Reduction of the CCD data 
has been performed in the standard manner using the 
ESO-MIDAS\footnote{MIDAS is 
developed and maintained by the European Southern Observatory.}
package. This includes dark and bias subtraction, flat-field correction
(we have used a mean of several dome flats taken in the appropriate
filter), sky subtraction. The cosmetic defects and projected stars
were excluded by median filtering (for small-scale defects and cosmic
rays) or by masking them with rectangular regions. The typical
uncertainty of the background level is (0.4--0.5)\%.
The $I$ band frames were obtained with insufficient exposure times for
the detailed photometry and we used the frames for the integral photometry only.

\begin{table}
\centering
\begin{tabular}{cccc}
\hline
\hline
 Data & Band-        & Exp &  $Z$     \\
      & pass         & (sec)& ($^{\rm o}$) \\
\hline                   
12/13 Aug 2002 & $B$ & 2$\times$900 & 35   \\
               & $V$ & 3$\times$600 & 34     \\
               & $R$ & 3$\times$600 & 38   \\
               & $I$ & 3$\times$600 & 35     \\	       
\hline
\end{tabular}
\caption{Observations at OPD}
\end{table}

The total magnitude of AM~1934-563 ($B=15.95\pm0.10$) found by us 
from the multi-aperture photometry is in good agreement with the 
NED\footnote{NASA/IPAC Extragalactic Database} ($B=15.97$) and 
LEDA\footnote{Lyon-Meudon Extragalactic Database} 
($B=15.93\pm0.10$) data. The two other galaxies show worse
agreement: 16.74$\pm$0.10 vs. 16.17$\pm$0.10 (LEDA) for the NW
companion, and 16.09$\pm$0.11 vs. 15.74$\pm$0.10 (LEDA) for
the southern galaxy of the triplet. Possible reasons for this
discrepancy are the contributions from nearby stars (PGC~400092 and
399718) and from the background galaxy (PGC~400092) (Figs.~1 and 2).
 
\subsection{Spectral observations}

The spectroscopic observations were performed with the 1.5-m 
telescope at CTIO in October 27, 2002, equipped with a Cassegrain 
spectrograph and CCD Loral 1K \#1, Arcon 3.9 IRAF software interface, 
square pixels size 15 micron. The grating \#35 of 600 lines/mm was 
centered at 675 nm (plate scale 18\farcs06/mm, scale slit 1\farcs33/pixel 
scale dispersion 1.498 \AA/pixel), and the spectral range 5900--7700 \AA.
The slit was set at 3$''$. The seeing during our observations was 
about 1$''$. A log of observations is given in Table~2. 

\begin{table}
\centering
\begin{tabular}{cccc}
\hline
\hline
 Data & Galaxy       & Exp &  P.A.($^{\rm o}$)     \\
\hline                   
27 \& 28 Sep 2002 & AM~1934-563 & 3$\times$1800$^s$ & 130$^{\rm o}$   \\
                   & AM~1934-563 & 1200$^s$ &  27$^{\rm o}$   \\
                   & PGC400092    & 2$\times$1200$^s$ & 130$^{\rm o}$   \\
                   & PGC399718    & 2$\times$1200$^s$ & 90$^{\rm o}$ \\	       
\hline
\end{tabular}
\caption{Observations at CTIO}
\end{table}

The reductions were carried out with standard techniques using 
IRAF\footnote{IRAF is distributed by NOAO, which is operated
by AURA Inc., under contract with the National Science Foundation.} 
and ESO-MIDAS packages. The emission-line kinematics was measured
by the Gaussian fit of the brightest emission lines present
in the spectra (H$\alpha$ and [NII]$\lambda$6583). We considered as
systemic velocities the values corresponding to the maxima of the
continuum intensity. Our derived heliocentric systemic velocity
for the AM~1934-563 (11649$\pm$10 km~s$^{-1}$) is in agreement with
previous optical measurements (see compilation of data in
van Driel et al. 2002) but significantly larger than the HI
value (van Driel et al. 2002). 

Table~3 summarizes the main observational characteristics of
three galaxies. The errors of heliocentric radial velocities
represent the formal standard deviations from the averaging
of velocities obtained using different lines and spectra.
Absolute magnitudes in the table were corrected for the internal 
absorption according to Tully et al.(1998).  

\begin{figure}
\centerline{\psfig{file=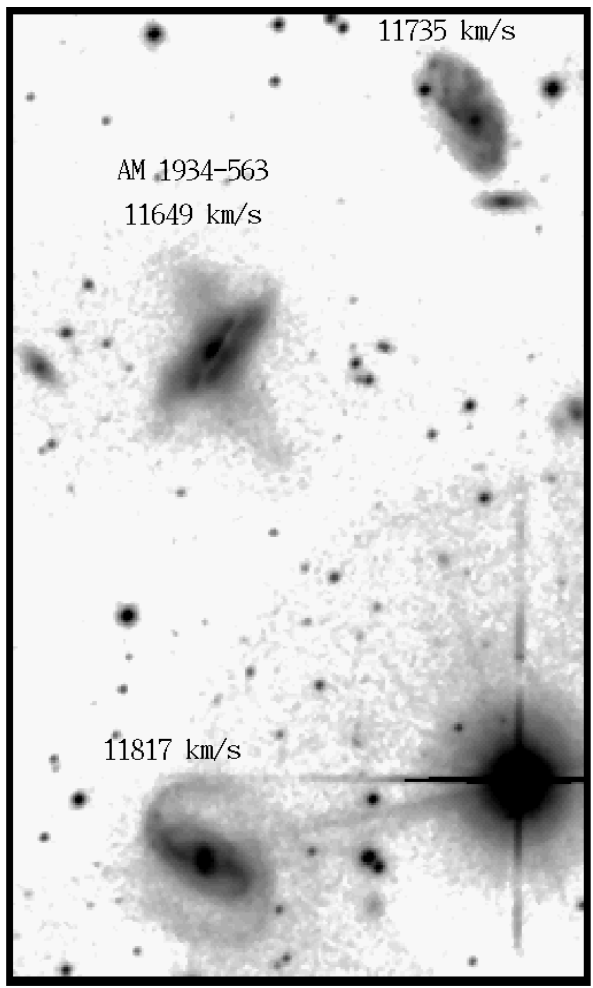,width=7.8cm,angle=0,clip=}}
\caption{$V$-band image of AM~1934-563. Image size is 2$'$.1 $\times$ 3$'$.6.
North is at the top, east to the left.}
\end{figure}

\begin{figure}
\centerline{\psfig{file=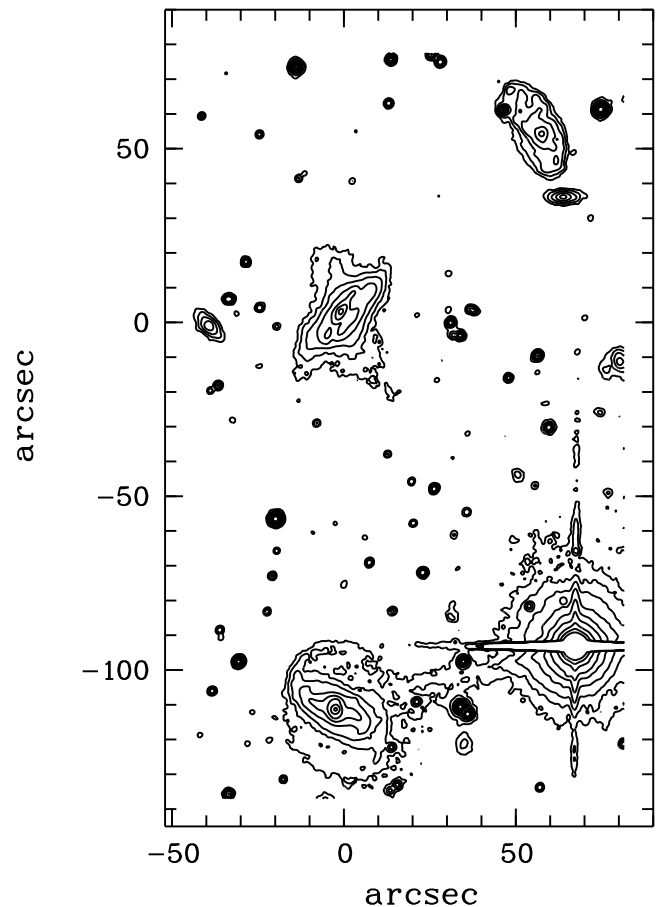,width=8.8cm,angle=-90,clip=}}
\caption{$V$-band contour map of AM~1934-563. The faintest contour
is 24.4 mag~arcsec$^{-2}$, isophotes step -- 0\fm75.}
\end{figure}

\begin{table*}
\begin{tabular}{lllll}
\hline
\hline
Parameter &  AM~1934-563 & PGC~400092 (NW)  & PGC~399718 (S) & Ref. \\
\hline
$\alpha(2000)$     & 19$^h$38$^m$38$^s$.52 & 19$^h$38$^m$31$^s$.55 &19$^h$38$^m$38$^s$.69                   \\
$\delta(2000)$     & --56$^{\rm o}$27$'$28$''$.5 &--56$^{\rm o}$26$'$36$''$.7 &--56$^{\rm o}$29$'$22$''$.3                 \\
Morphological type             & SBa/b:    & Sd/Irr:   & SBc: &       \\
Heliocentric systemic velocity$^*$ (km~s$^{-1}$)& 11649$\pm$10 & 11735$\pm$6&11817$\pm$26 \\
Adopted luminosity distance$^{**}$  &            & 167.6 Mpc & &   \\
Scale                        &  & 0.75 kpc/1$''$    &\\
Major axis, D$_{26}$~($\mu_B=26$) & 50$''$ (38 kpc)& 36$''$ (27 kpc) & 43$''$ (32 kpc) &               \\
Axial ratio                  &0.45: & 0.40& 0.6:&             \\ \\
Total apparent                 &          & &&\\
magnitudes and colors:         &          & &&\\
$B_{\rm T}$         & 15.95$\pm$0.10 &  16.74$\pm$0.10 & 16.09$\pm$0.11 \\
$(B-V)_{\rm T}$     & +0.94$\pm$0.04 &  +0.61$\pm$0.04 & +1.15$\pm$0.05 \\
$(V-R)_{\rm T}$     & +0.60$\pm$0.03 &  +0.38$\pm$0.03 & +0.45$\pm$0.05 \\
$(R-I)_{\rm T}$     & +0.51$\pm$0.1 &   +0.2$\pm$0.1   & +0.35$\pm$0.1 \\
$(J-H)_{\rm 2MASS}$            & +0.83 & & +0.57 & [1]       \\
$(H-K)_{\rm 2MASS}$            & +0.44 & & +0.29 & [1]       \\
Galactic absorption ($B$-band) & 0.22  & 0.22    & 0.22 & [2] \\
$K$-correction ($B$-band)      & 0.07  & 0.01    & 0.05 & [3] \\
Absolute magnitude, $M_B^0$    & --21.1 & --20.2 & --20.7 & \\
$L_{FIR}$ ($L_{\sun}$)        & 8.8 $\times$ 10$^{10}$ & & & NED \\
SFR$_{FIR}$ (M$_{\sun}$/yr)   & 45  & & & \\
\hline \\
\end{tabular} 

$^*$ -- conventional radial velocity obtained as $cz$ (Fairall 1992);

$^{**}$ -- $\Omega_m$=0.3, $\Omega_{\Lambda}$=0.7, $H_0$=70 km~s$^{-1}$/Mpc;

[1] -- Skrutskie et al. 1997;
[2] -- Schlegel et al. 1998;
[3] -- Bicker et al. 2004

\caption{General properties of the galaxies}
\end{table*}

\section{Results and discussion}

\subsection{Dynamical characteristics of the triplet}

AM~1934-563 and two neighboring galaxies show very close 
radial velocities (Fig.~1, Table~3) and very probably form
physical triplet.
Considering this group as an individual dynamical system, we 
computed the following standard quantities (see Karachentseva \&
Karachentsev 2000 for the corresponding formulae):

\begin{itemize}
\item The dispersion of galaxy velocities with respect to the center:
$s_v$ = 69 km~s$^{-1}$.

\item The mean harmonic projected distance: $r_H$ = 82.2 kpc.

\item The dimensionless crossing time of the system:
$\tau$ = 0.17 (in units of the Hubble time, $H_0^{-1}$).

\item The virial mass: M$_{vir}$ = 1.3 $\times$ 10$^{12}$~M$_{\sun}$.

\item The virial mass-to-luminosity ratio for the triplet in solar
units: $f$ = 14.  
\end{itemize}

From the comparison of the above characteristics with the
corresponding median values of known isolated triplets
(Table~3 in Karachentseva \& Karachentsev 2000) we can conclude that
the AM~1934-564 system shows about twice smaller $s_v$ value.
The characteristic size of the triplet -- $r_H$ -- is close to median 
values for the southern and northern triplets. The relatively low 
value of the observed dispersion leads to a large crossing time, and 
small virial mass and mass-to-luminosity ratio in comparison with 
typical triplets. 

\subsection{Nuclear spectra of the galaxies}

\begin{figure}
\centerline{\psfig{file=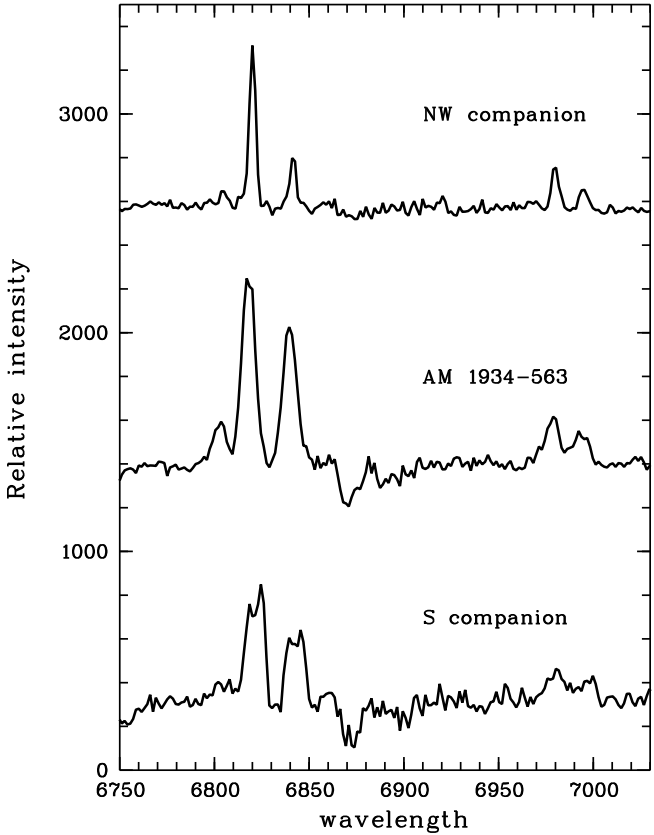,width=8.8cm,angle=-90,clip=}}
\caption{Arbitrary scaled nuclear H$\alpha$ spectra of the triplet members.}
\end{figure}

The nuclear H$\alpha$ spectra of galaxies are presented in Fig.~3. 
The results of our measurements of the nuclear emission-line properties 
within $4'' \times 3''$ are summarized in Table~4.

\begin{table}
\centering
\begin{tabular}{cccc}
\hline 
\hline
 Parameter & AM~1934-563 & NW galaxy &  S galaxy  \\
\hline                   
{\scriptsize FWHM(H$\alpha$), km~s$^{-1}$} & 320$\pm$20 & $<$100  & 440$\pm$40  \\
{\scriptsize W$_{\lambda}$(H$\alpha$), \AA} & 12.1$\pm$0.7 & 17.5$\pm$0.6 & 9.1$\pm$0.8    \\
{\scriptsize $\rm [NII]\lambda$6583/H$\alpha$} & 0.74$\pm$0.04 & 0.32$\pm$0.08 & 0.6$\pm$0.1   \\
{\scriptsize $\rm [SII]/H\alpha$}  & 0.33$\pm$0.1 & 0.45$\pm$0.1 & 0.25$\pm$0.1  \\
\hline 
\end{tabular}
\caption{Nuclear emission-line properties}
\end{table}

The NW member of the triplet shows typical HII region-like spectrum. 
AM~1934-563 demonstrates relatively large
[NII]$\lambda$6583/H$\alpha$ ratio and noticeable width of the
emission lines; but the galaxy shows comparatively faint
[OI]$\lambda$6300 line ([OI]$\lambda$6300/H$\alpha \leq 0.05$).
Therefore, we can classify its spectrum as a transition 
-- AGN/HII -- type (Veilleux \& Osterbrock 1987). 

Emission lines in the nuclear spectrum of PGC~399718
(S galaxy) are double-peaked and wide (Fig.~3), reflecting very
steep rotation curve visible in the two-dimensional spectrum.
(The atlas by Ho et al. (1995) gives several examples of such
objects -- e.g., NGC~488, NGC~3245, and NGC~4772.)
The amplitude of splitting determined from the Gauss decomposition
of the H$\alpha$ and [NII]$\lambda$6583 contours is
250--280 km~s$^{-1}$. 

\subsection{AM~1934-563}

\subsubsection{Main galaxy}

The main object is a distorted edge-on spiral galaxy (Figs.~1 and 2).
The outer parts of the galaxy show significant warp of the isophotes
with amplitude of $\sim8^{\rm o}$. The bright inner part of the disk with P.A.=148$^{\rm o}$ is crossed by a prominent dust lane and is inclined with respect to the outer isophotes by about $13^{\rm o}$. This inner inclined structure could represent an almost edge-on bar. Detailed inspection shows that
the dust lane is split and embraces the galaxy nucleus from SE and NW. 

Fig.~4 displays the surface brightness profiles of AM~1934-563
approximately along the major axis and along the bright inner
part of the outer structure (see Sect.~3.3.2). The surface brightness
was averaged within 5$\times$5 pixels or 0\farcs9$\times$0\farcs9.
The major axis profile
shows central peak and two depressions at $\mid r \mid \approx 5''$
probably related to the dust lanes. At $\mid r \mid \geq 10''$
the major axis profile can be approximated by an exponential
disk with a central surface brightness $\mu_0(B)=20.06 \pm 0.19$
(corrected for the Milky Way extinction and $K$-correction) and 
exponential scale-length $h(B)=5\farcs1 \pm 0\farcs3$ (3.8 kpc).
Therefore, the disk of AM~1934-563 is typical for high surface 
brightness, star-forming disks of other known PRGs with spiral hosts 
(Reshetnikov 2004). 

The disk of AM~1934-563 shows strong color gradients (Fig.~4b):
central parts of the galaxy are red ($B-V \approx +1.1$,
$V-R \approx +0.9$), while the outer ones are blue 
($B-V \approx +0.6$, $V-R \approx +0.3$). The observed ratios
of the scale-lengths in different color bands are 
$h(B)/h(V)=1.18\pm0.11$, $h(B)/h(R)=1.25\pm0.12$. 
Such color gradients are very common among face-on (e.g., de Jong 1996) 
and edge-on (e.g., de Grijs 1998) spiral galaxies. De Grijs (1998)
argues that the observed radial color gradients in edge-on
galaxies largely represent their dust content.   

\begin{figure*}
\centerline{\psfig{file=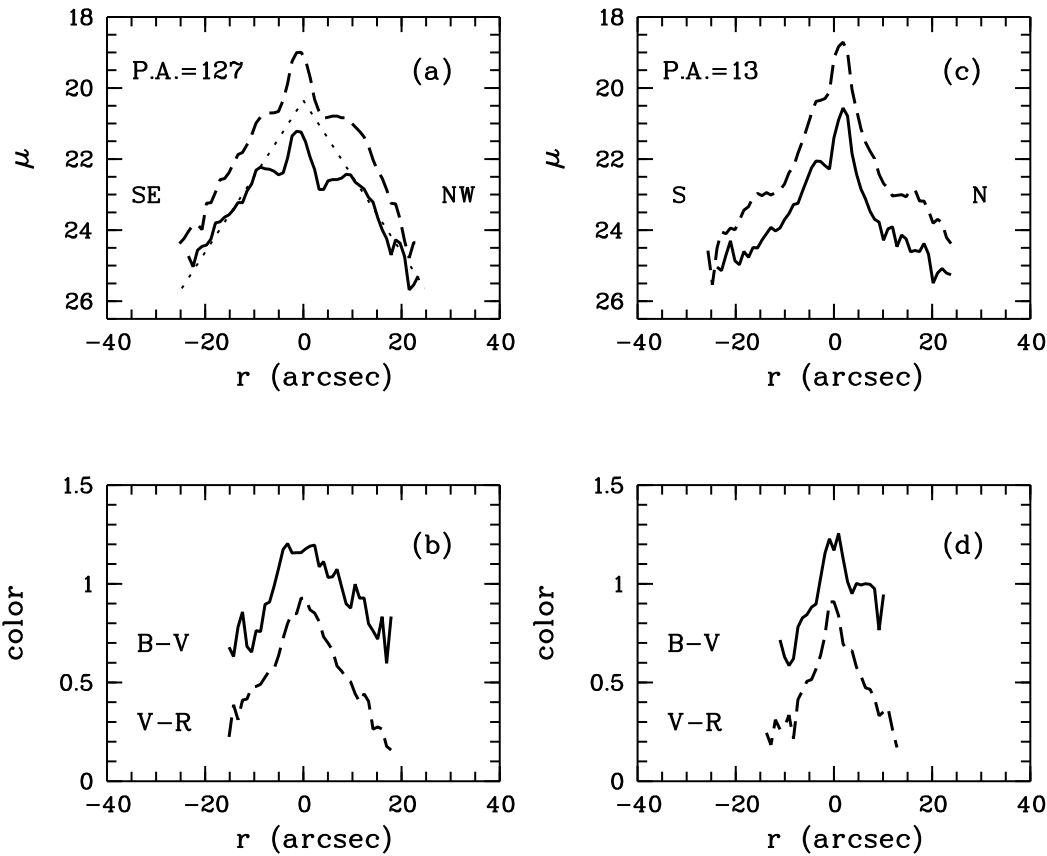,width=14cm,angle=-90,clip=}}
\caption{Photometric profiles for AM~1934-563: {\bf a), b)} along
the apparent major axis; {\bf c), d)} along the major axis of
the suspected ring. Solid line in {\bf a)} and {\bf c)} represent
the distributions in the $B$ passband, dashed lines in $R$.
Dotted line in {\bf a)} shows the symmetric exponential model
for the disk.}
\end{figure*}

The optical colors of the galaxy corrected for the Milky Way
absorption, $K$-correction and rather uncertain internal absorption
are $(B-V)_0\approx0.6-0.7$ and $(V-R)_0\approx0.4-0.5$. 
These values are usual for the Sa--Sb galaxies (Buta et al. 1994;
Buta \& Williams 1995).

AM~1934-563 is a strong source of far-infrared emission (PRC, Table~3).
Converting the far-infrared luminosity to the star formation rate
(Bushouse 1987), we obtain 45~M$_{\sun}$/yr and $L_{FIR}/L_B$=3.4. 
Both estimates are very high in comparison with the data for isolated and 
even for interacting galaxies (e.g., Bushouse 1987). We suppose that
the observed IRAS flux can be contaminated by the neighboring 
members of the triplet. Even in the case where we
divide the SFR and $L_{FIR}/L_B$ estimates by 3 (roughly), AM~1934-563
is a galaxy with very active star formation.

\begin{figure}
\centerline{\psfig{file=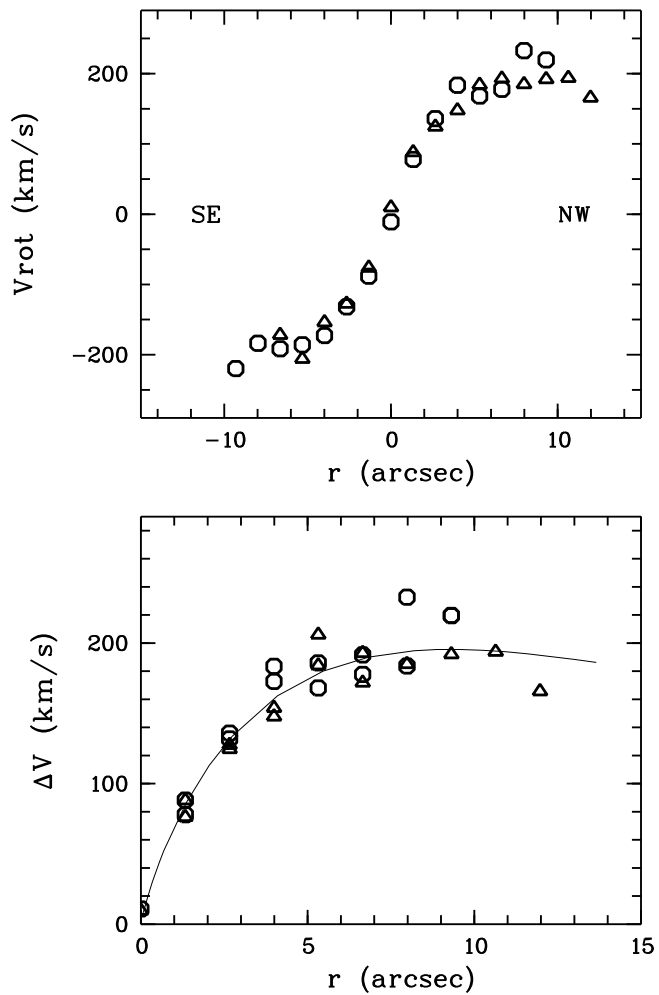,width=8.8cm,angle=-90,clip=}}
\caption{Top -- radial velocities distribution along the major axis of
AM~1934-563 (P.A.=130$^{\rm o}$), 
bottom -- averaged rotation curve (circles -- H$\alpha$
data, triangles -- [NII]$\lambda$6583). Thin solid line represents 
the rotation curve of an exponential disk with scalelength of 4\farcs1 
(scalelength value in the $R$ filter).}
\end{figure}

Fig.~5 shows the emission-line rotation curve of the galaxy (after
correction for the cosmological stretch). Thin solid line in
the bottom panel is a fit by an exponential disk with
$h(R)=4\farcs1$ (3.1 kpc) and intrinsic axial ratio 0.2 (Monnet \&
Simien 1977). It
can be seen that the exponential disk approximation gives a good
description of the observed rotation curve within 12$''$ (9 kpc) from the
nucleus. The maximum rotation velocity of the galaxy obtained 
is V$_{max}$=195 km~s$^{-1}$. The actual V$_{max}$
value can be somewhat larger due to peculiar appearance of the 
galaxy and uncertain inclination. The Tully--Fisher
relation predicts that the extinction-corrected luminosity of
a galaxy with V$_{max}$=195 km~s$^{-1}$ is $M(B)\approx-21$ (Tully et al. 1998,
Kannappan et al. 2002). Therefore, AM~1934-563 equatorial disk lies close to 
the Tully--Fisher relation for normal spirals (at least, in the
first order approximation). 

Assuming a spherical mass distribution and a flat rotation curve
within the optical radius (R$_{26}$=19 kpc), we can estimate
the AM~1934-563 mass as 1.68 $\times$ 10$^{11}$~M$_{\sun}$. 
Therefore, the mass-to-luminosity ratio in the $B$ passband is 4 
in solar units -- a rather normal value for a giant spiral galaxy. 

\subsubsection{Possible polar ring}

\begin{figure}
\centerline{\psfig{file=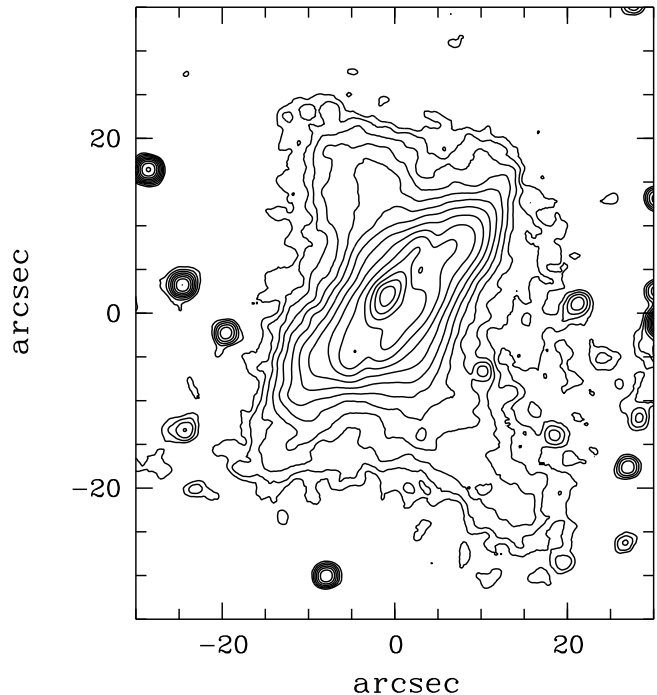,width=8.8cm,angle=-90,clip=}}
\caption{Contour map of AM~1934-563 from the combined $B+V+R$ image.
Isophotes are separated by factor 1.5.}
\end{figure}

Fig.~6 presents a deep contour map of AM~1934-563. The most fascinating
feature of the galaxy is a giant (optical diameter reaches 60$''$
or 45 kpc) inclined S-shaped structure which crosses the main
body. From the morphological point of view, the off-plane structure
resembles the polar rings or disks of the polar-ring galaxies (see 
examples in the PRC).
The appearance of AM~1934-563 is very similar to NGC~660 (see Fig.~1 in 
van Driel et al. 1995). But, in contrast to NGC~660 which is a
sub-$L^*$ galaxy, AM~1934-563 is a giant spiral galaxy. 

The ring is very faint. Even the brightest inner parts show $\mu(B) \approx 24$
(Fig.~4c). As the main galaxy, the ring shows a color gradient
(Fig.~4d).

The polar ring (or polar disk) shows a strong warp. Its inner part is inclined
about 57$^{\rm o}$ with respect to the AM~1934-563 major axis and 
the outer one is inclined about 72$^{\rm o}$. 

The ring is asymmetric.
It extends up to 26$''$ to the NE from the nucleus, and up to 34$''$
in the opposite direction.
The color distribution along the ring is also asymmetric: the
NE (shorter) part of the ring looks redder in the $B-V$ (Fig.~4d) and
this color asymmetry can be traced up to the end of the ring. 

The observed colors of the ring are relatively blue in comparison with
the colors of central galaxy. Its mean colors corrected for
the Milky Way extinction and $K$-correction are $B-V=0.7$,
$V-R=0.35$ (ring colors were measured over 6\farcs5 diameter
apertures placed at 15$''$--25$''$ on the both sides of the nucleus of the
galaxy). Taking into account that the ring is almost edge-on,
its real "face-on" colors may be significantly bluer. 

The total observational luminosity of the ring is
about (10--15)\% of AM~1934-563. Therefore, the rings absolute
luminosity can reach $M(B) \approx -18.5...-19$. So the AM~1934-563 
ring is comparable with the brightest ring structures among 
nearby PRGs (e.g., Reshetnikov et al. 1994). 

\begin{figure}
\centerline{\psfig{file=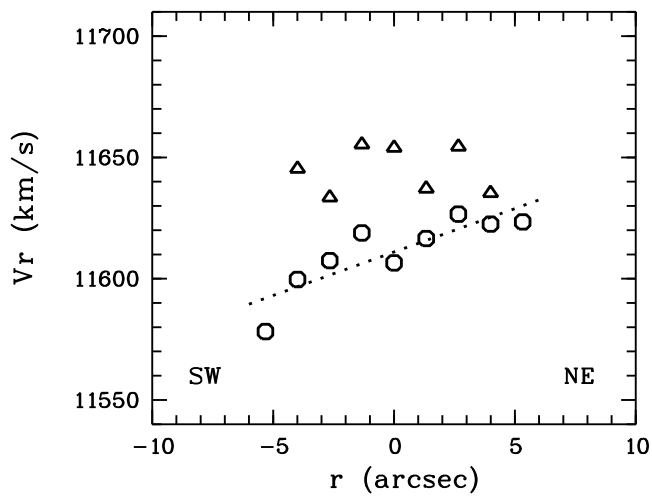,width=8.8cm,angle=-90,clip=}}
\caption{Radial velocities distribution along the major axis of
the ring of AM~1934-563 (circles -- H$\alpha$ data, triangles -- 
[NII]$\lambda$6583).}
\end{figure}

Fig.~7 shows the distribution of radial velocities along 
P.A.=27$^{\rm o}$. Unfortunately, the slit goes through the faint 
periphery regions of the ring but misses the bright inner ones. The 
forbidden [NII]$\lambda$6583 line demonstrates almost
constant velocity close to the galaxy systemic velocity.
The H$\alpha$ line gives only marginal evidence for the velocity 
gradient along the ring (the dotted line in Fig.~7 shows linear
fit of the data). Interestingly, the direction of this gradient
coincides with the direction of the main body rotation (Fig.~5).

Along P.A.=27$^{\rm o}$ we observe the rise of the [NII]/H$\alpha$
ratio from 0.7--0.8 in the nucleus to $\sim$1.3 at $\mid r \mid =
3''-4''$. Similar increases in the [NII]/H$\alpha$ ratio have 
been observed recently in a number of edge-on spirals, e.g., UGC~10043
(Matthews \& de Grijs 2004 and references therein). Shock-heating
due to starburst-driven wind is the usual explanation of such line ratios.
In the case of AM~1934-563 we need additional spectral observations
to confirm the presence of a large-scale wind. It is interesting to
note that two other PRGs with young or forming rings are classified as
 ''superwind'' galaxies -- NGC~660 (Armus et al. 1990)  and
NGC~6286 (Shalyapina et al. 2004). Possible polar-ring related
spiral UGC~10043 also shows the presence of a large-scale wind
(Matthews \& de Grijs 2004). One can propose that nuclear
starburst and, correspondingly, starburst-driven wind extending up 
to the outer kinematically decoupled structure are the natural consequences
of an external accretion event, as suggested by numerical models (BC03, 
and the model in this paper).

A systematic shift in radial velocities derived from H$\alpha$ and
[NII] lines is evident in Fig.~7. It is not an unique feature
since many spirals show the same behavior (e.g., Afanasiev et al. 2001,
Moiseev 2002). The difference in observed velocities may be
explained by the simple assumption that the [NII] emission line 
originates partially in shocks, 
due to a wind or a bar (Afanasiev et al. 2001).

\subsection{Companion galaxies}

Both companion galaxies are giant spirals. The NW galaxy (PGC~400092)
demonstrates asymmetric, slightly irregular morphology without
pronounced spiral arms. The optical colors of the galaxy (corrected
for the Milky Way absorption and $K$-correction) are rather
blue ($B-V=0.55$, $V-R=0.35$) and are typical for a
late-type or irregular galaxy. A small edge-on (may be background) galaxy
with $B=19.5$, $B-V=0.9$, $V-R=+0.6$ (observational values)
is located 19$''$ SW of the PGC~400092 nucleus.

The southern galaxy of the triplet (PGC~399718) possess a bar and 
a two-arms spiral structure. The $B-V$ color of the galaxy
looks unusually red for spirals and this can be an indication of
a high inclination. A high velocity gradient
at P.A.=90$^{\rm o}$ (see sect.~3.2) also supports this point of view. 
Therefore, the galaxy structure may be not planar - with
almost edge-on inner part and more face-on outer ones. 

\section{Modeling of the AM~1934-563 morphology}

\subsection{Numerical techniques}

The simulations were performed with a $N$-body particle
code which includes gas dynamics and star formation (see BC03).  
The main characteristics of this code is that the potential is computed 
by FFT on a 3D Cartesian grid. The size of the grid is 128$^3$, and 
the Fourier images are suppressed by the method of James (1977). 
The softening length is equal to the size of a grid cell, which is 
of the order of 1 to 3 kpc, according to the models. The particles, 
stars, gas and dark matter, are in total number between 2 and 6$\cdot$10$^5$. 
The gas dynamics is modeled by the sticky-particles scheme.

\subsection{Results}

There are too many free parameters in a numerical model
to fit the  morphology of AM~1934-563  in details (see discussion
in BC03). Therefore, we decided to find
a solution which describes reasonably only the general observational  
properties of the galaxy. We have considered two scenarios for
the polar ring formation: 1) minor merger with a small 
companion, where the ring is formed by the tidal disruption of the 
smallest galaxy and 2) tidal accretion of matter from a massive, 
gas rich donor galaxy to the polar ring (as in BC03). The major merger 
scenario (see Introduction) has not been simulated, but models of 
this scenario can be found in Bekki (1998) and BC03.

\paragraph{Minor merger}
In the framework of the first scenario, we have considered a minor 
merger with an inclined orbit, where the ring results from the 
disruption of the companion. In the present case, the mass of the 
companion should be 10--20\% of the mass of the host galaxy: more 
massive companions would disturb the host galaxy too much and 
form a S0-like object (Bournaud et al. 2004, 2005a), less massive ones can 
only form rings that are much fainter than in the observed system. 
On the basis of our observations, the main galaxy stellar mass
was set to M$_{stars}=2 \times 10^{11}$~M$_{\sun}$. 
The gas fraction was fixed as M$_{gas}$/M$_{stars}$=0.2.
We assumed a spherical dark matter halo, described by a Plummer 
sphere of scale-length 30 kpc and truncation radius 60 kpc, 
with a dark-to-visible mass ratio M$_{DH}$/M$_{stars}$=0.6 inside 
the stellar radius. This mass was chosen to fit the rotation curve 
amplitude. The characteristic size of the exponential disk is taken 
from the surface brightness fit. 

\begin{table*}
\centering
\begin{tabular}{lccccccc}
\hline
\hline
$N^o$&$V_{\infty}$ (km/s)& $\Theta$ ($^{\rm o}$) & R$_p$ (kpc) & & rotations & tilt & $R_1/R_2$ \\
\hline
1 & 250 & 25 & 80  & & 1.4 & 40 & 2.5 \\
2 & 100 & 25 & 50  & & 0.5 & -- & 3.5 \\
3 & 250 & 25 & 50  & & 1.3 & 30 & 2.2 \\
4 & 100 & 25 & 80  & & 0.8 & 25 & 1.9 \\
5 & 150 & 25 & 120 & & 0.9 & 20 & 2.2 \\
6 & 150 & 25 & 40  & & 0.6 & -- & 2.5 \\
7 & 150 &  0 & 80  & & 0.8 & 0  & 2.0 \\
8 & 150 & 35 & 80  & & 0.8 & 35 & 2.3 \\
9 & 250 & 35 & 80  & & 1.2 & 45 & 2.8 \\
\hline
\end{tabular}
\caption{Run parameters and results of simulations of minor mergers, where the 
disruption of the companion forms a ring-like feature. We indicate the orbital 
parameters (column 2--4): relative velocity, inclination of the orbital plane w.r.t. 
a polar plane, and impact parameter. We also give estimates of three characteristics 
of the ring, showing its unclosed aspect: the number of rotations visible in the 
ring-like structure, the tilt of the ring plane between the two sides of the ring 
(related to the precession of the companion orbit), and the ratio of the extend of 
both sides of the ring (related to the orbital decay of the companion). The tilt 
of the ring orientation could not be estimated in two cases, because the ring-like 
structure was not complete enough. These values have been estimated about 300 Myr 
after the merging of the companion.}
\end{table*}

\begin{figure*}
\centerline{\psfig{file=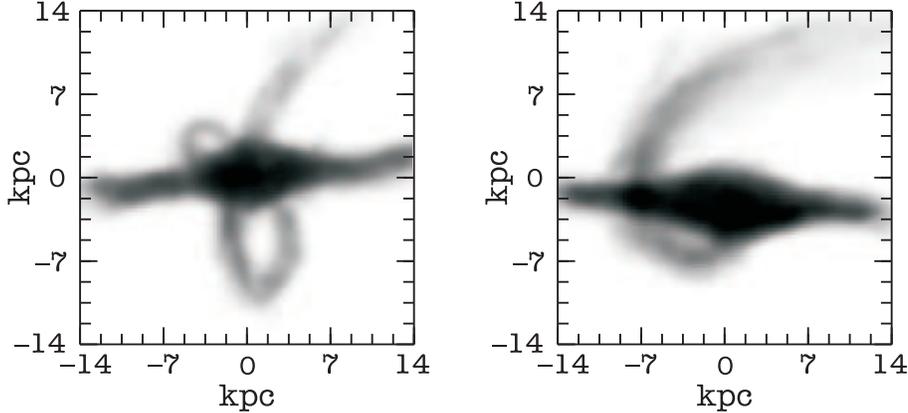,width=12cm,clip=}}
\caption{Projection of the polar ring in two simulations of minor mergers, 
seen 300~Myr after the merging of the companion (left: run1, initial orbital 
plane seen close to edge-on -- right: run2, initial orbital plane seen face-on). 
The three aspect making the ring ''unclosed'' are illustrated by these figures: 
the orbital plane is largely tilted in one rotation (left), the ring makes less 
than one rotation (right), and there is a high decay in the radius between 
both sides of the ring (left and right). Quantitative results for more simulations 
are given in Table~5. We used a non-linear scale adjusted to enhance the visibility 
of the ring-like feature.}
\label{fig8}
\end{figure*}

We have simulated several minor mergers with a companion mass of 15\% the primary 
mass. Such a mass ratio is required to form a massive enough ring, without disturbing 
the host disk more than in the observed case. The internal characteristics of the 
companion are fixed as M$_{gas}$/M$_{stars}$=0.25 and M$_{DH}$/M$_{stars+gas}$=1.5, 
and orbital parameters given in Table~5. The result of two representative simulations 
is shown on Fig.~\ref{fig8}. We found that a minor merger with such a mass ratio 
can only form {\it unclosed} ring-like inclined features:
\begin{itemize}
\item for a ring inclined by at least 25 degrees from the polar axis (as in the observed 
system), differential precession has affected the orbit of the progenitor companion, 
which results in a significant tilt between both sides of the ring (visible when the ring 
is seen close to edge-on, e.g. on Fig.~\ref{fig8})
\item debris from the companion can often be detected on less than one rotation , making 
the ring incomplete, in particular when the impact parameter or relative velocity of 
the companion are small. This is mainly visible when the ring is seen face-on, but for 
edge-on projections its results in a large discrepancy in the luminosity of both sides 
of the ring (one side of the projected ring can be several times more luminous 
than the other one)
\item because the orbital decay of the companion, the two sides of the ring have 
largely different sizes (one side is generally more extended than the other one by a 
factor of 2--3).
\end{itemize}
These three aspects are described quantitatively in Table~5 for several simulations of 
minor mergers. The two sides of the ring of AM~1934-563 have comparable sizes, 
comparable masses, and form a continuous structure with a rather constant orientation. 
This cannot be reproduced by simulations of minor mergers, for an inclined ring. 
In particular the discrepancies in the mass/extent of both sides of the observed 
ring are much lower than in simulations of minor mergers. The general problem of 
this scenario is that a companion on an inclined orbit does not have enough 
time to stretch more or less uniformly along the orbit before merging with the primary, 
because the time-scale of dynamical  friction (that has a typical time-scale of 0.55~Gyr in 
our simulations) is not much larger than the dynamical time-scale (that has an average 
value of 0.4~Gyr). Thus, differential precession and radial decay affect the orbit faster 
than the material is captured into the ring, which does not allow the morphology of 
a closed ring to be reproduced. It thus seems unlikely that the observed polar ring 
is the result of a minor merger. Moreover, these unclosed ring-like structures are 
generally short live, and are dispersed by differential precession in less than 1~Gyr. 
With a more massive companion, they could be longer-lived, because self-gravity 
would reduce the effects of differential precession, but such a companion would make 
the host galaxy become an extremely early-type disk galaxy, or even an S0 
(see Bournaud et al. 2004, 2005a), which is not observed. 

We have not varied the dark halo density profile. This could be an important parameter, 
because it changes the timescale for dynamical friction, which can affect the ring being 
closed or not. However, we have largely varied the velocity and impact parameter of the 
companion. Then, the disruption of the companion occurs at various radii, around 
various densities of dark matter, thus, under different degrees of dynamical friction. 
This change does not allow a closed ring to be formed. Changing the halo profile 
would change the dynamical friction over a rather similar range, so the conclusion 
would be the same, namely that it cannot explain the formation of the ring of AM~1934-563.

We finally conclude that the observed polar ring is certainly not the result of the 
stripping of a small companion, during a minor merger.

\paragraph{Tidal accretion}
In a second scenario we have considered an unbound encounter of the
primary (target) with a gas rich giant spiral galaxy (donor). Tidal mass
transfer from the donor to the target is then assumed to form a polar
ring around the target. The mass of the target galaxy is 
$2 \times 10^{11}$~M$_{\sun}$. In the best fitting model that we have found, 
the donor galaxy has a stellar disk of radius 17 kpc, and its gaseous disk
extends up to 42 kpc. Its visible mass is $3.6\times10^{11}$~M$_{\sun}$, and
the mass fraction of gas among its visible matter, including gas outside
the stellar radius, is 0.33. The target galaxy has a stellar radius of 14 kpc. 
It contains a gas fraction M$_{\mathrm{gas}}$/M$_{\mathrm{stars}}$=0.18, with no gas outside its 
stellar disk, as is the case for many early-type galaxies. Both galaxies 
are embedded in a
dark halo described by a Plummer sphere of radial scale-length 30 kpc and
truncation radius 60 kpc. The dark-to-visible mass ratio inside the
stellar disk radius is 0.6 for the target galaxy, and 0.5 for the donor
galaxy. The initial conditions of the encounter have to be 
selected carefully, in order to favor inclined accretion of the tidal 
tail. The suitable conditions we have found are: impact parameter 130 kpc, 
relative velocity of V= 145 km~s$^{-1}$, almost perpendicular orientation 
of the donor disk and the orbital plane with respect to the host 
galaxy disk -- but not exactly perpendicular, so that the ring is 
both inclined and warped. The simulation produced an inclined, large-scale, 
faint stellar ring around the main galaxy. If we denote $t=0$ the 
beginning of the simulation, the pericenter passage occurs at $t=780$ Myr. 
A tidal bridge is formed between the two galaxies, and is accreted to a 
polar structure (Fig.~9). The polar ring is first unclosed and irregular, 
but becomes closed and regular in a few $10^8$ yr, matching the 
observed appearance of AM~1934-563. While rings resulting from
minor-merging companions appear to remain unclosed in our simulations, the
ring is here unclosed only during a short time, when it is an unrelaxed,
still forming, structure. Even less than 1 Gyr after its formation the ring
is likely to be observed as a closed structure. Thus, that the ring of 
AM~1934-563 is closed is not a strong constraint for the scenario of 
tidal accretion from a massive donor. The reason is that the ring material 
is captured by the host galaxy over about 0.3 Gyr, faster than the dynamical 
friction timescale, which allows the ring to be closed. This is the contrary 
of the minor merger scenario, where the material is captured over more than 
one rotation of the companion, i.e. at least 0.5 Gyr: in this other case we 
have shown that the ring does not get closed.

Fig.~10 presents contour plots of stars in the simulated polar-ring galaxy 
at $t=1600$ Myrs (820 Myrs after the pericenter passage). At that time 
the donor galaxy has flown away at about 250 kpc distance (which of 
course, can appear much closer in sky projection) and does not perturb 
the target any more. As for the polar ring, it has become more regular 
than on the earlier stages shown in Fig.~9, even if still asymmetrical. 
This model reproduces the main characteristics of AM~1934-563: 
\begin{itemize}
\item the polar ring inclination, and the relative sizes of the polar 
ring and the host galaxy, are well reproduced. The projected distribution 
of stars along the major axis of the model galaxy is well approximated 
by an exponential disk with $h=2.8$ kpc (for AM~1934-563 we have found 
$h=3.1$ kpc).
\item the model fits the warp of the polar ring, and the S-shape of the 
host galaxy. The latter is caused by a small warp but also to a 
barred-spiral structure, since the host disk is not observed perfectly 
edge-on
\item the asymmetrical aspect of the ring, with its southern part more 
extended and less luminous than the northern one, is reproduced
\item in the model, the contribution of the ring to the total luminosity 
is about 11\%. This percentage agrees reasonably with the observational 
value (Sect.~3.3.2): the mass of the polar structure is comparable to 
the mass of the observed polar ring.
\end{itemize}

\begin{figure*}
\centerline{\psfig{file=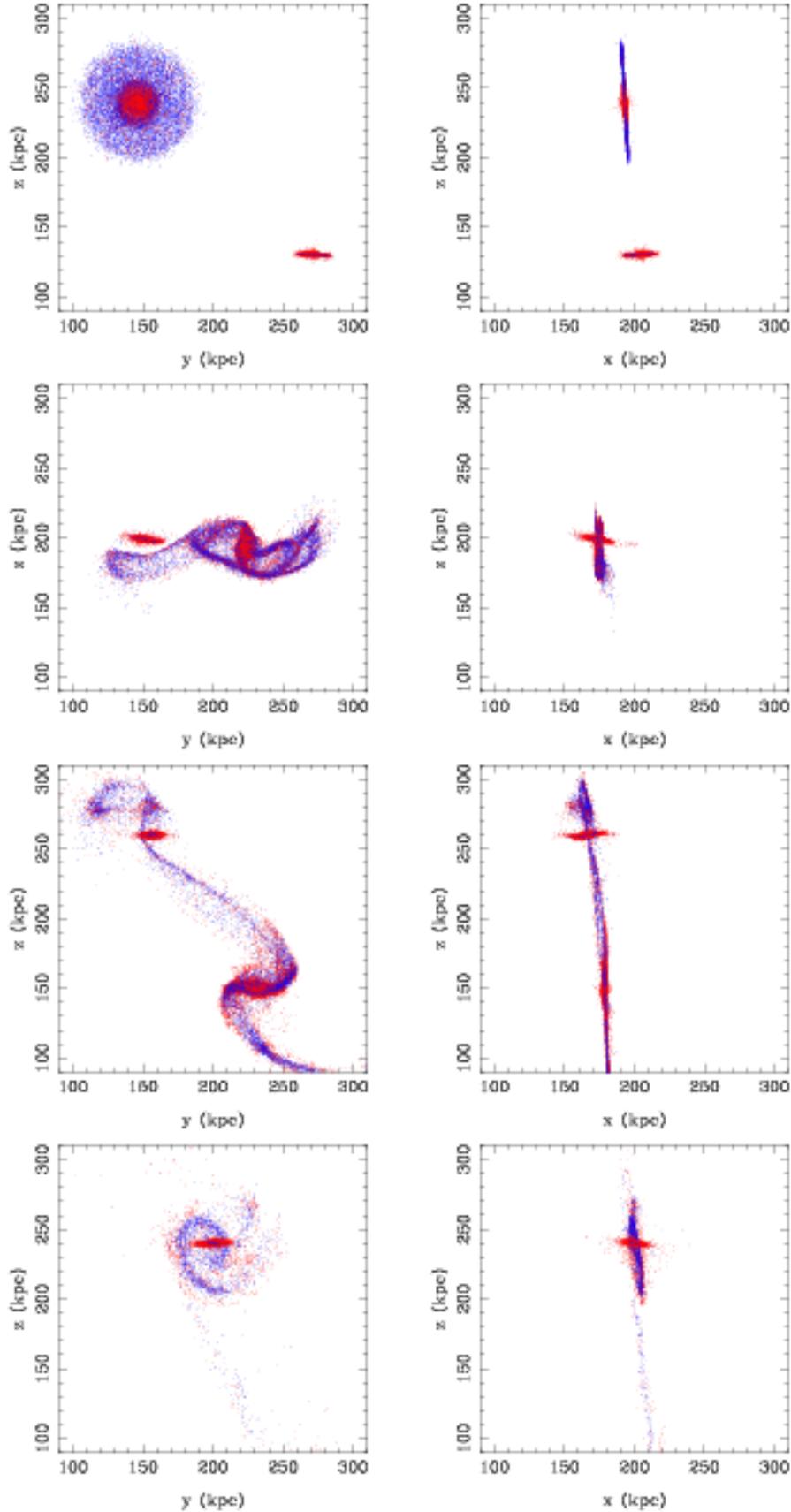,width=12cm,clip=}}
\caption{Two projections of the polar-ring formation run (two columns), 
seen at $t=$200, 700, 900, and 1150 Myr. Stellar particles are in red, 
gas clouds in blue (30000 particles are plotted). In the last snapshot, 
the ring has just formed and is not closed yet. It is closed and more 
regular a few $10^8$ Myr later, as shown in Fig.~10, while closed rings 
are not formed in our simulations of disruption of a minor companion.}
\end{figure*}

\begin{figure}
\centerline{\psfig{file=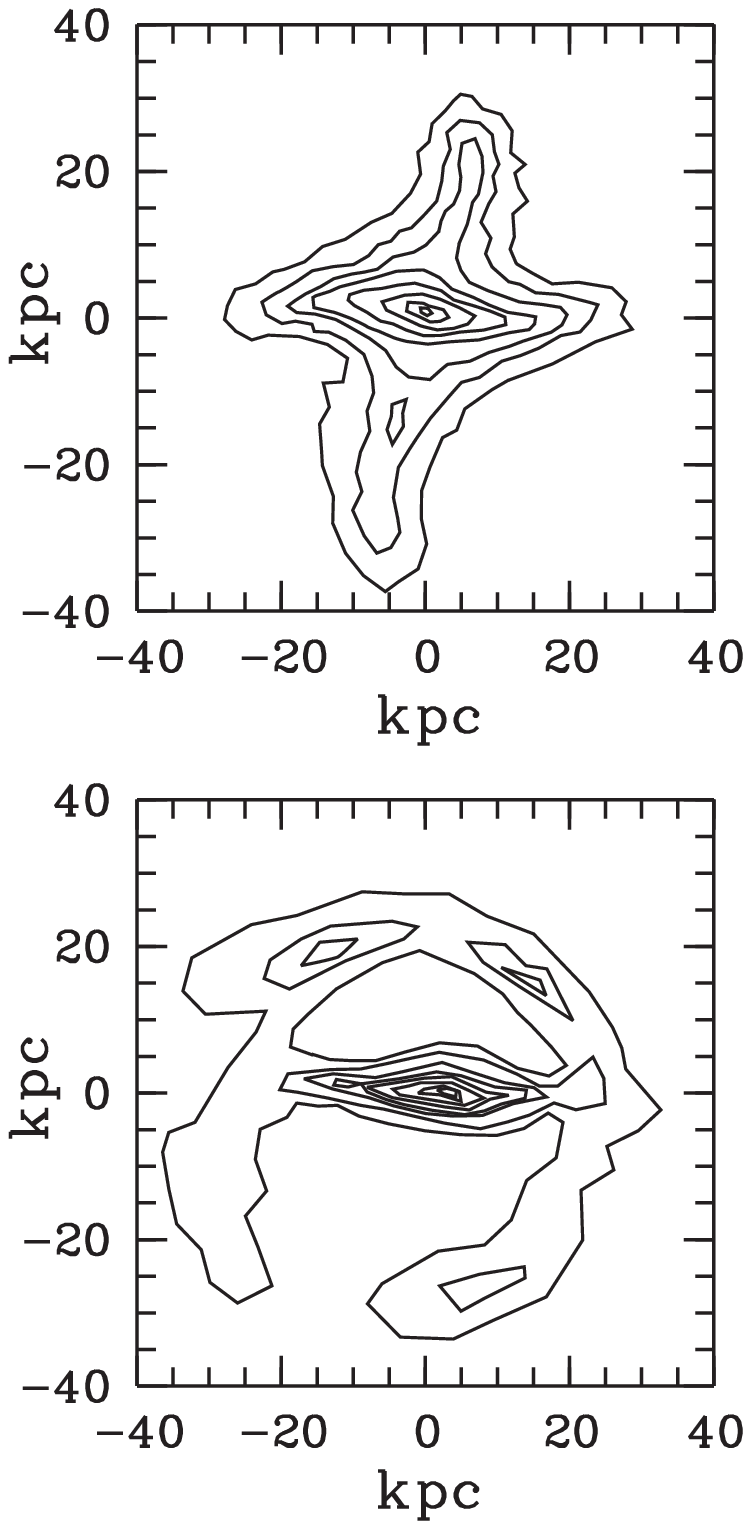,width=8cm, clip=}}
\caption{Two projections of the stellar distribution in the simulated 
polar-ring galaxy (820 Myrs after the pericenter passage). The projection 
above (with the host galaxy seen nearly edge-on) has been chosen to 
fit the main characteristics of the observed system, as detailed in the 
text. The projection below, with the host disk seen edge-on and the polar 
ring face-on, is nearly orthogonal to the previous one. The contours 
are logarithmic, separated by a factor 2. The ring has rapidly become a
closed structure, contrary to what happens in the minor merger scenario,
and the mass distribution along the ring is still asymmetrical, which the major 
merger scenario fails to reproduce (see text).} 
\end{figure}

Let us note that the matter transferred from the donor to the 
target galaxy is not only gas. As can be seen in Fig.~9, some stars 
from the donor are also captured in the polar disk, without being 
dispersed, but most of them have been formed just before the ring, inside 
the tidal bridge. That some stars are formed before the ring may lead 
to an over-estimation of the age of the ring, yet these stars do not 
dominate the mass and are not much older than the ring, so that this 
over-estimation may not exceed 1--2$\times 10^8$ yr. A small fraction of 
dark matter is also accreted from the donnor galaxy, but for our model with 
non-rotating dark haloes, this does not exceed 5\% of the donnor dark mass.

\subsection{Formation scenarios}

At least four scenarios can be proposed to account for young, massive 
polar rings surrounding pre-existing host galaxies:

\begin{itemize}

\item the major merger scenario (e.g., Bekki 1998), where the two 
merging protagonists have resulted one in the host, and the other in 
the polar ring: the polar ring is here the remnant of a pre-existing disk

\item the disruption of a small companion on a polar orbit, the debris 
of which form the polar ring

\item tidal transfer of material from a massive donor during a galaxy 
encounter. The donor is not necessarily merging with the host galaxy, 
and keeps its identity after the event

\item accretion of gas infalling from cosmic filaments towards the 
polar structure. Gas can be accreted gradually and form a purely 
gaseous a polar ring in a first step, then an external perturbation 
triggers star formation in the ring, as proposed by Cox et al. (2001) 
for the system on interacting galaxies II~Zw~70/71.

\end{itemize}

The last three ones can be regarded as three subtypes of the accretion 
scenario. Let us note that the first and third of these variants do 
not predict the presence of companions around PRGs, which is supported 
by observations. For example, Brocca et al. (1997) find that 
statistically, the environments of PRGs show no excess of close 
companions, which is consistent with the majority of PRGs forming 
via either long-term secular gas accretion or via mergers in which 
the companion is destroyed. However, this does not rule out accretion from a 
gas-rich donor, for the donor may have left the scene. Iodice et al (2002a,b) 
favor the merger scenario of two disk galaxies, from the study of stellar 
populations and morphological structure of the host; however, the constraints 
are not strong enough, since in any scenario, gas can be accreted by the host 
and induce star formation, while the host is perturbed, so that a large range 
of stellar ages should coexist in the perturbed host. The presence or not of a 
diffuse envelope of stars around the PRG is more constraining (see BC03).

The major merger scenario has been shown to be less likely than the tidal 
accretion scenario, especially for inclined rings (BC03). Moreover, the 
asymmetric ring of AM~1934-563 cannot be well fitted by this scenario. 
Indeed, this scenario assumes that the host galaxy as merged with a 
''victim'' disk galaxy that has given birth to the polar ring. Since the 
southern part of the ring in AM~1934-563 is more extended than the northern 
one, the center of this ``victim'' galaxy should be  south of the host 
galaxy, which should make the southern part of the polar ring brighter 
that the northern one. On the contrary, the northern part of the ring 
is brighter than the southern one. We have checked in simulations of 
BC03 that the more extended part of the ring is at least as luminous as 
the shortest part, when the ring is formed in a major merger.

We have explained above that the disruption of a small companion, during 
a minor merger, is unlikely to have formed the polar ring of AM~1934-563, 
for the ring would certainly appear unclosed, because the orbit of the 
companion is unclosed itself. Moreover, forming a massive enough ring, 
without disturbing the host disk too much, is a strong constraint on the 
mass of this companion.

On the contrary, the accretion scenario invoking a tidal mass transfer from 
a massive donor, without galaxy merger, has been shown to succeed in 
reproducing the characteristics of AM~1934-563. Furthermore, the small 
velocity dispersion of this group (see Sect.~3.1) could result from past 
galaxy interactions without merger, during which dynamical friction has 
dissipated the galaxy relative velocities. The velocity dispersion would 
be significantly higher if the three galaxies came from large distances 
and had not interacted yet. The small observed velocity dispersion 
also suggest that galaxies are close to their apocenter, confirming 
they have been closer, grazing, in the past.

Which of the two other triplet members is the donor galaxy? Both galaxies 
are late-type and large enough to supply the matter to form the ring in 
AM~1934-563. The southern galaxy (PGC~399718) appears a more promising 
candidate, since it is perturbed and shows non-planar structures. Also, 
its red $B-V$ color can be a consequence of gas stripping process during 
the interaction with AM~1934-563. 

To confirm and to make more precise our model, new HI observations of 
the triplet 
are required, to search for a possible bridge between AM~1963-563 and 
the donor, and to estimate the HI content of the galaxies and of the ring. 
The extended southern part of the ring could be the origin of a bridge, 
as suggested by our numerical model (see the snapshot at $t=1150$ Myr in 
Fig.~9). Also 2D kinematical data for AM~1934-563 are highly needed 
both to get kinematical constraint on the formation scenario.

\section{Conclusions}

From our reported photometry and spectroscopic study of the AM~1934-563 
system, we have studied a confirmed polar-ring galaxy (PRG). We have 
tested by numerical simulations the various possible scenarios, minor 
mergers, or tidal accretion of matter. It appears that the minor merger 
model does not reproduce the observed parameters in AM~1963-563. The tidal 
accretion of matter succeeds to reproduce the overall morphology, including 
the inclination of the ``polar'' ring, its mass, and the S-shapes of 
the ring and the host disk. The small relative velocities of the 
three galaxies of the triplet may indicate that tidal interactions 
have occurred recently.

The case of accretion is quite likely for AM~1963-563, since it belongs 
to a triplet, with two possible donor galaxies. This is also the case 
for the prototype polar ring galaxy, NGC~4650A, which belongs to a dense 
group. The donor may also have left the group and be today further away 
from the PRG, because it can have a large relative velocity in this 
scenario. The small observed velocity dispersion in the group of AM~1963-563 
may only mean that we are seeing this flattened group face-on, 
so that the donor could have left the scene.

The present numerical simulations have shown that the exchanged 
matter between the donor and the target galaxy is not composed 
exclusively of gas, but a small amount of stars are also accreted in 
the polar ring. This means that the age of the polar ring formation, 
estimated from the age of its stellar population, might be over-estimated, 
i.e. polar rings could be slightly younger than they appear. Yet, most 
of these accreted stars are found to be formed inside the tidal bridge 
that leads to the ring formation, so the over-estimation of the age 
of a polar ring may probably not exceed 1--2$\times10^8$ yr.

We also found that the host galaxy lies close to the Tully-Fisher relation. 
This is the case for most host galaxies of polar rings, and had led 
Iodice et al. (2003) to the conclusion that PRGs contain dark matter 
along their polar structure. Unfortunately, the present observations 
did not enable us to study the ring kinematics in detail, thus we could 
not get constraints on its dark matter content. However, this PRG is a 
good target to study the distribution of dark matter, since the radius 
of the host galaxy and the polar ring are similar. Indeed, the too 
small radius of the host galaxy is an important problem in many PRGs. 
New HI or 2D-kinematical observations are required to study the 
distribution of dark matter.

In our simulations of the accretion scenario, we have assumed that gas is 
tidally accreted from a gas-rich donor. A variant of this scenario has been 
suggested by Cox et al. (2001): gas could form a ring via gradual accretion, 
and the stellar counterpart forms when a perturbation induces star formation. 
At this stage, we do not try to discriminate between these two possibilities: 
we only want to know whether the ring is formed by gas accretion in general, 
or by galaxy interaction/mergers.

Then, the case of AM~1934-563 (like other PRGs according to BC03) appear to support 
the view that galaxies are not formed in one event only, but they acquire mass 
all along their lives, as is also attested by morphological properties of disk 
galaxies, like bars (Bournaud \& Combes 2002), and lopsidedness (Bournaud et al. 2005b). 
Material accreted during the life of galaxies is quite visible in the case of PRGs 
because of its peculiar perpendicular orientation. In most general case, the material 
will be accreted in the plane, or with some angle lower than about 60$^{\rm o}$ 
from the plane, which can give rise to warps, disk lopsidedness of the disk, or 
density waves like arms or bars.

\acknowledgements{This work has been progressing while VR was visiting
LERMA in Paris, thanks to a CNRS 3-months visitor grant.
VR acknowledges support from the Russian Foundation 
for Basic Research (03-02-17152) and from the Russian Federal Program 
''Astronomy'' (40.022.1.1.1101). 
M.F.-A. and M. de O.-A. thank the partial
support of the Funda\c{c}\~{a}o de Amparo \~{a} Pesquisa do Estado de Minas
Gerais (FAPEMIG) and the Minist\'{e}rio da Ci\^{e}ncia e Tecnologia (MCT,
Brazil). We would like to thank the anonymous referee whose detailed
remarks have helped improve the paper. The computations in this work 
have been carried out on the Fujitsu NEC-SX5 of the CNRS computing center, 
at IDRIS. This research has made use of the NASA/IPAC Extragalactic 
Database (NED) which is operated by the Jet Propulsion Laboratory, 
California Institute of Technology, under contract with the National 
Aeronautics and Space Administration. We made use of the LEDA database 
(http://leda.univ-lyon1.fr). }


\end{document}